\pdfoutput=1
\documentclass[fleqn,usenatbib]{mnras}

\usepackage{newtxtext,newtxmath}

\usepackage[T1]{fontenc}

\DeclareRobustCommand{\VAN}[3]{#2}
\let\VANthebibliography\thebibliography
\def\thebibliography{\DeclareRobustCommand{\VAN}[3]{##3}\VANthebibliography}

\usepackage{graphicx}	
\usepackage{amsmath}	
\usepackage{subcaption}

\title[Fossil records]{An early interaction between the Sagittarius dwarf and the Milky Way I: Fossil signatures in the phase-space structure of the disc}

\author[P. Das et al.]{
Payel Das,$^{1}$\thanks{E-mail: p.das@surrey.ac.uk}
Jason A. S. Hunt$^{1}$
\\
$^{1}$University of Surrey, Stag Hill, Guildford, Surrey, GU2 7XH, United Kingdom\\
}

\date{Accepted XXX. Received YYY; in original form ZZZ}

\pubyear{\the\year{}}

\begin{document}
\label{firstpage}
\pagerange{\pageref{firstpage}--\pageref{lastpage}}
\maketitle

\begin{abstract}
Several independent lines of evidence point to an early encounter between the Milky Way and the Sagittarius dwarf galaxy $5$--$7$~Gyr ago. We present four controlled $N$-body+hydro simulations, one of an isolated host galaxy and three of a single interaction between a Milky-Way-like host and a Sagittarius-like satellite on an unbound polar orbit. Relative to the fiducial satellite (with total mass $0.7\times10^{11}$~M$_\odot$), one run increases the satellite's mass by a factor of $1.5$ and another its crossing speed by a factor of $1.4$, giving pericentric speeds of $466$--$634$~km~s$^{-1}$ at pericentric radii of $R\simeq15$--$18$~kpc. Comparing the present-day, phase-space structure in radial bins by age of newly-formed thin-disc stars across the runs, we find that stars that existed at the time of the passage are thickened by up to $\sim37$~per cent in mass-enclosed height and heated by up to $\sim5$~km~s$^{-1}$ in vertical velocity dispersion relative to the control, while stars formed after the passage are unaffected. The heating signature is also evident when comparing the age-vertical velocity dispersion relation against a heating relation with a power law of $0.38$, obtained by fitting the isolated host. A single, disc crossing from an early interaction thus suffices to leave lasting, datable fossils in the age-resolved vertical structure of a galactic disc. The fiducial run's dispersion jump of $\sim4$~km~s$^{-1}$ matches best with that observed in the outer low-$\alpha$ disc of the Milky Way \citep{das+24}, corresponding to a perturber mass of $0.7\times10^{11}$~M$_\odot$, with a velocity of $466$~km~s$^{-1}$ at a pericentric radius of $16.7$~kpc, though the inversion from jump amplitude to perturber mass is good only to a factor of $\sim1.5$.
\end{abstract}




\section{Introduction}
\label{sec:introduction}

Galaxies like our Milky Way host significant satellite systems. For host galaxies of masses $\sim10^{12}M_{\odot}$, there should be $\sim$ ten $10^9$ satellites, three $10^{10}$ satellites, and one $10^{11}$ satellite \citep{read+19,santos+22}. Some satellites are massive enough that as they undergo pericentric passages with the host galaxy and possibly eventually merge, they should not only deposit debris and generate streams in the stellar halo \citep{johnston+96,helmi+99,bell+08,naidu+20,horta+23a}, they should also dynamically perturb the disc of the host. 

Phase-space impacts induced by satellite encounters are often transient. \citet{toomretoomre72} showed, using a restricted three-body technique with a non self-gravitating disc that close tidal encounters excite bridges and tails. This was later extended using fully self-gravitating $N$-body(+SPH) simulations, in which the disc, halo, and (where included) gas respond self-consistently to their own gravity. This showed that even non-merging flyby encounters excite long-lived, self-amplified grand-design spiral structure \citep{pettitt+16,semczuk+17}. This response is not confined to the disc plane: \citet{kumar+22} show that flybys simultaneously excite vertical breathing and bending motions \citep[e.g.][]{Widrow+14}, coupling the in-plane tidal spiral to a vertical phase-space response analogous to what drives the Milky Way's phase spiral \citep{antoja+18,hunt+21}. 

Satellite encounters can also permanently alter disc structure and kinematics. \citet{tothostriker92} used analytic arguments to show that a disc's observed thinness limits the rate of satellite infall, and \citet{quinn+93} showed that significant heating occurs once a satellite's orbital kinetic energy matches the disc's vertical kinetic energy. \citet{villalobos_helmi08,villalobos_helmi09} accreted two-component satellites onto a thin disc and, using a $5{:}1$ mass-ratio merger, showed that these encounters leave lasting phase-space substructure in the resulting thick disc. \citet{kazantzidis+08} showed using $N$-body simulations that satellite accretion produces persistent morphological signatures such as flaring, thickening, bars, warps. \citet{martig+14a,martig+14b} used cosmological disc-galaxy simulations to show that, once dissected into mono-age populations, both a disc's structural properties and its age-velocity dispersion relation retain imprints of its accretion and heating history. 

\begin{table*}
    \centering
    \begin{tabular}{|c|c|c|c|c|c|c|c|c|c|c|}
    \hline
      Component & $N_{\mathrm{p}}$ & $M$ ($M_\odot$) & c & $r_{\mathrm{s}}$ (kpc) & $R_{\mathrm{d}}$ (kpc) & $r_{\mathrm{h}}$ (kpc) & $\sigma_{\mathrm{R}}/\sigma_z$ & $T$ (K)  \\
      \hline
       DM halo & $2\times10^6$ & $6.08\times10^{11}$ & 9 & - & - & - & - & -  \\
       Stellar bulge & $2.4\times10^5$ & $1.2\times10^{10}$ & - & 0.3 & - & - & - & -   \\
       Thin disc & $9\times10^5$ & $4.5\times10^{10}$ & - & - & 3.0 & 0.25 & 2.5 & - \\
       Thick disc & $1\times10^5$ & $0.5\times10^{10}$ & - & - & 2.0 & 0.9 & 4 & -  \\
       Gas disc & $8.9\times10^5$ & $0.44\times10^{10}$ & - & - & 7.0 & 0.1 & - & $10^4$  \\
       Gas halo & $1\times10^5$ & $0.5\times10^{10}$ & - & - & - & - & - & $10^7$ \\
       \hline
       Satellite DM halo & $1\times10^4$ & $7.08\times10^{10}$ & 20.5 & - & - & - & - & - \\
       Satellite stars & $2\times10^3$ & $1.41\times10^8$ & - & - & 0.7 & 0.32 & 3.0 & - \\
       Satellite gas & $2.0\times10^3$ & $9.53\times10^8$ & - & - & 2.0 & 0.8 & - & $10^4$ \\
       \hline       
       Chemistry (dex) \\
       \hline
       Component & $\Delta$[Fe/H]/R & [Fe/H]$_0$ & $\sigma$([Fe/H]) & $\Delta$[Fe/H]/Gyr & $\Delta$[$\alpha$/Fe]/R & [$\alpha$/H]$_0$ & $\sigma$([$\alpha$/Fe]) & $\tau$ (Gyr) & $Z/Z_{\odot}$ & [$\alpha$/Fe] \\
       \hline   
       Stellar bulge & 0.0 & 0.0 & 0.0 & 0.0 & 0.0 & 0.3 & 0.0 & 6-7 & - & -\\
       Thin disc & -0.05 & 0.2 & 0.02 & -0.04 & -0.16 & 0.0 & 0.02 & 0-3 & - & -\\
       Thick disc & 0.0 & -0.4 & 0.1 & 0.0 & 0.0 & 0.3 & 0.2 & 3-6 & - & -\\
       Gas disc & -0.05 & 0.2 & 0.02 & - & 0.0 & 0.0 & 0.02 & - & - & -\\
       Gas halo & - & - & - & - & - & - & - & - & 0.03 & 0.3\\
       \hline
       Satellite stars & -0.01 & -0.4 & 0.1 & 0.0 & 0.0 & 0.3 & 0.2 & 0-5 & - & - \\
       Satellite gas & -0.05 & -0.2 & 0.02 & - & -0.05 & 0.1 & 0.02 & - & - & - \\
    \end{tabular}
    \caption{Particle numbers, masses, structural properties and chemical properties of the initial conditions for the Milky-Way-like host and the fiducial satellite (used in the fiducial simulation \texttt{pmr6}). The `massive satellite' used in \texttt{pmr6\_massive}\ has identical parameters except for a DM halo mass $1.5$ times larger. All particle numbers and masses quoted are the realized totals. Three components are smoothly truncated at large radii relative to the analytic profiles from which they are sampled: the host and satellite DM haloes, whose underlying NFW profiles (to which the concentrations refer) have virial masses of $0.9\times10^{12}$~M$_\odot$ and $10^{11}$ ($1.5\times10^{11}$)~M$_\odot$ respectively, and the host gas disc, which retains $887{,}273$ of $10^{6}$ sampled particles. All other components -- the stellar discs, bulge, hot gas halo, and the satellite's stars and gas -- are realized in full. In the chemistry block, radial gradients are per kpc, the age gradient per Gyr, $\sigma$ denotes the Gaussian scatter, and $\tau$ is the stellar age range assigned to each painted population.}
    \label{tab:sim_params}
\end{table*}

There is also direct evidence that pericentric passages and mergers leave a permanent imprint on a galaxy's distribution function in observational data. In the Milky Way, the discovery of the Gaia-Enceladus/Sausage showed that an ancient, relatively massive merger permanently reshaped the local velocity distribution and possibly dynamically heated the pre-existing disc into the thick disc \citep{belokurov+18,helmi+18,das+20}, with independent age-dating of the event confirming a single merger event that quenched star formation in the satellite \citep{gallart+19,myeong+19,das+20}. Outside the Local Group, orbit-superposition dynamical modelling of integral-field data in external early-type and lenticular galaxies has begun to identify kinematically and chemically distinct orbital families attributable to past accretion events \citep{poci+19,poci+21,zhu+20,zhu+22}.

Evidence has accumulated for an early Sagittarius--Milky Way encounter some $5$--$7$~Gyr ago, well before the more recent pericentric passages usually invoked to explain present-day disc substructure. Orbit modelling of the Sagittarius stream dates the first disc-crossing infall to $\sim5$~Gyr ago \citep{law_majewski10,purcell+11,laporte+18}. Independently, Sagittarius' own star-formation history shows typical dwarf-galaxy star formation until $5$--$7$~Gyr ago, after which it quenches abruptly \citep{de_boer+15}, and its core is dominated by an intermediate-age population formed $5$--$6$~Gyr ago \citep{siegel+07}. These scenarios are both consistent with gas stripping at this first infall. From the perspective of the Milky Way, \cite{das+24} find a sharp jump in the vertical velocity dispersion of the outer low-$\alpha$ disc that they interpret as a signature of the first pericentric passage of Sagittarius. \citet{ruiz_lara+20} independently find evidence for a significant starburst around the Solar Neighbourhood of the Milky Way, $5.7$~Gyr ago, using Gaia colour-magnitude-diagram fitting. \citet{mackereth+19} also find that mono-age disc populations in the Milky Way undergo a significant change in velocity structure at an age of $6$--$8$~Gyr. 

\begin{figure*}
    \centering
    \includegraphics[width=\linewidth]{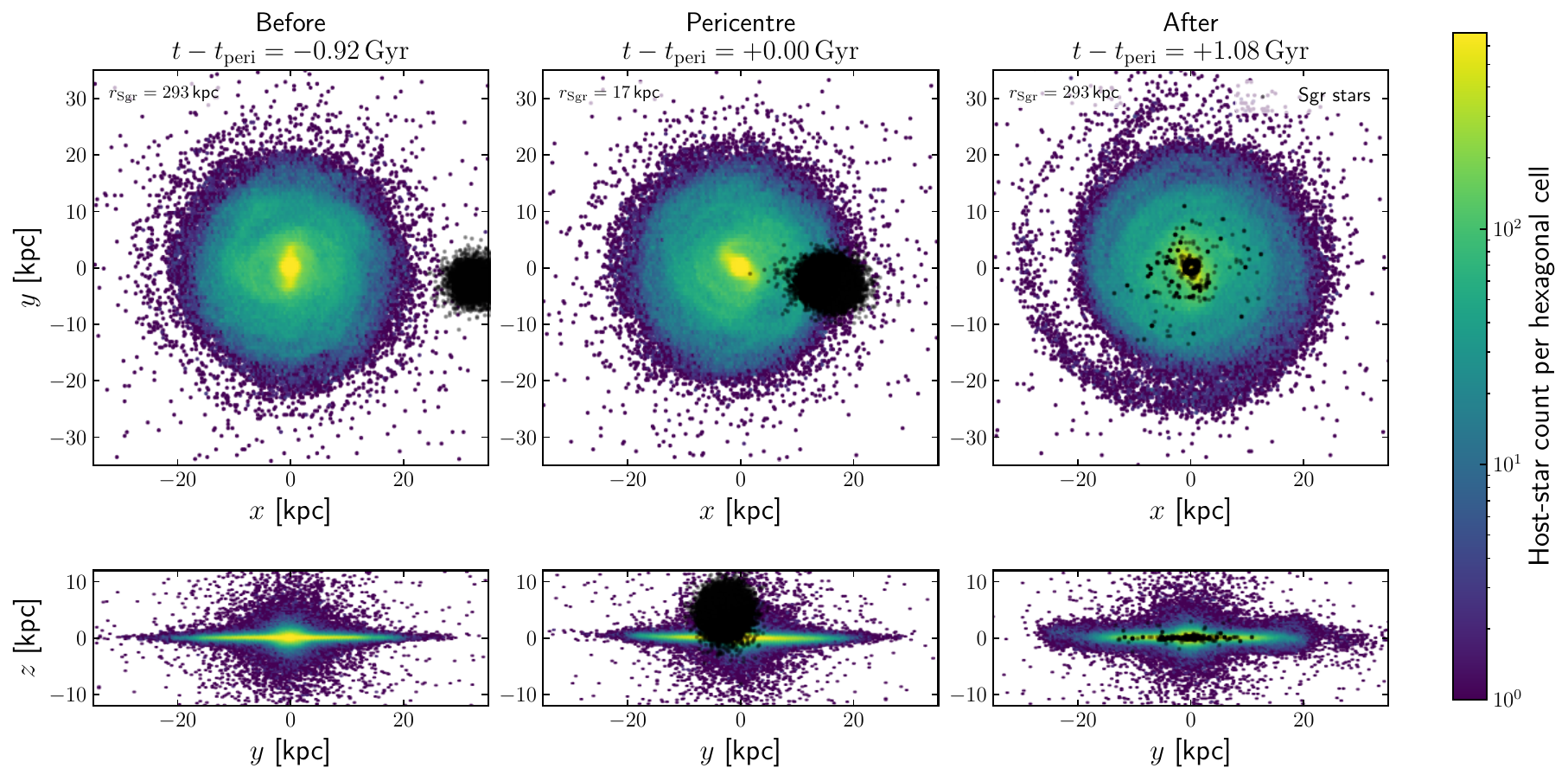}
    \caption{Face-on ($x$--$y$, top) and edge-on ($y$--$z$, bottom) projections of the Milky-Way-like disc in \texttt{pmr6}\ around the pericentric passage, which occurs $7.1$~Gyr before the final snapshot (the present day). Colour shows the host star count per hexagonal cell on a logarithmic scale; black points show satellite-origin stars. One Gyr after pericentre, the satellite has been disrupted into a stream threading the disc plane, and the host retains a pronounced thickened vertical profile. Each panel is annotated with the three-dimensional separation $r_{\rm Sgr}$.}
    \label{fig:before_during_after}
\end{figure*}

In this paper, we run four controlled $N$-body+hydro simulations of an interaction between a Milky-Way-like galaxy and a Sagittarius-like dwarf galaxy to see if an early encounter can explain some of the features seen in \cite{das+24}.  We discuss the set up of the simulations in Section \ref{sec:simulation_setup}, the orbits and mass loss of the Sagittarius-like dwarf galaxy in Section \ref{sec:orbit_mass_loss}, the impacts of the interactions on the disc phase-space structure in Section \ref{sec:sat_impacts}, then end with the Discussion in Section \ref{sec:discussion} and Conclusions in Section \ref{sec:conclusions}.

\section{Simulation setup}\label{sec:simulation_setup}
The simulations here are designed as an idealised experiment on whether a single satellite crossing can explain the disc chemodynamical features in \cite{das+24}. As such, we do not attempt to generate a realistic analog of the Milky Way--Sagittarius merger, but rather to have a reasonably quiet and stable disc model which we can `hit' with a satellite. First, we set up the initial conditions for our Milky-Way-like (MW) galaxy. We use the initial condition generator \texttt{pgen}\ as described in \cite{Kawata+14}, following the method of \cite{Springel+05}, to construct a system consisting of thin and thick stellar discs, a stellar bulge, a cold gas disc, a hot gas halo, and a live dark matter halo. 

The stellar particles in the initial condition are `painted' with initial age and metallicity information to approximate the Milky Way's chemical evolution with the oldest stars in the bulge, intermediate ages in the thick disc, and the youngest in the initial thin disc, with the $\alpha$ and iron trends to match. Each population is assigned a linear trend of [Fe/H] and [$\alpha$/Fe] with radius and with age, plus a Gaussian scatter; the coefficients are listed in the chemistry block of Table~\ref{tab:sim_params}. We then generate a satellite perturber using the same initial condition generator. The dynamical and chemical parameters of both the host and the satellite initial conditions can be found in Table \ref{tab:sim_params}. The resulting particle masses are $3.0\times10^{5}$~M$_\odot$ for the host dark matter, $5\times10^{4}$~M$_\odot$ for the painted stars and hot halo gas, $5\times10^{3}$~M$_\odot$ for the disc gas, and $7.1\times10^{6}$, $7.1\times10^{4}$ and $4.7\times10^{5}$~M$_\odot$ for the satellite dark matter, stars and gas respectively. We place our satellite on the host's polar axis at an initial separation of $800$~kpc ($1519$~kpc in \texttt{pmr6\_faster}), moving inward nearly radially at $239$~km~s$^{-1}$ ($500$~km~s$^{-1}$), with a small in-plane offset chosen so that it crosses the disc at $\sim17$~kpc, roughly following the estimated disc crossing location from \cite{das+24}. 

One caveat here is that we make no attempt to reproduce the early cosmological accretion history of the Milky Way, and while our painted gradients are an approximation of that evolution, those stars would be inherently biased by our painting scheme, and only stars which form self-consistently from the gas are included in the subsequent analysis. The models are also relatively low resolution by current standards, but we do not require fine phase-space resolution in this work, and they are appropriate to analyse disc-wide trends.

The other main caveat is that we choose to initialise the satellite on an unbound, impulsive orbit. The justification of this is twofold: Firstly, while the exact properties of the first encounter of the Sagittarius dwarf galaxy are not well known, some prior works \citep[e.g.][]{hunt+21} find that the initial pericentre in their model is impulsive, while the later passages are not. Secondly, in such a regime, it is also found that there is a significant gap between the first and second pericentre \citep[e.g. $\sim2-2.5$ Gyr in][]{laporte+18,hunt+21} such that the satellite has a large apocentre at that stage of the merger, and very little influence on the host between the first and second encounters. The purpose of this paper is to explore the effect of that first, potentially impulsive encounter, rather than the later merger, so it suffices to have the dwarf have no further influence on the host disc after the first crossing. Section~\ref{sec:orbit} quantifies how impulsive the realized encounters actually are, and for which parts of the disc.

The simulations are evolved for 10 Gyr using the $N$-body / SPH galaxy simulation code \texttt{ChaNGa}\rm\ \citep{Jetley:2008,Jetley:2010,Menon:2015}, with self-consistent star formation, feedback and chemical evolution tracking iron and oxygen mass fractions. The simulations are run with a master timestep of $1$~Myr, with 10 rungs for individual timestepping, and a gravitational softening of 20 pc. Star formation occurs with a minimum density of $10\ \mathrm{cm}^{-3}$, and maximum temperature of 1000 K. The simulations have metal cooling and diffusion \citep{Shen+10} and the stellar feedback energy is set at $10^{51}$~erg per supernova, implemented with the `blastwave' prescription of \cite{Stinson+06}. Snapshots are written every $10$~Myr, giving the $1000$ snapshots used for the orbit and mass-loss analysis of Section~\ref{sec:orbit_mass_loss}. The present-day phase-space and chemical analyses of Section~\ref{sec:sat_impacts} use the final snapshot only.

We run four simulations in total. \texttt{pmr6}\ is the fiducial MW+Sgr run. \texttt{pmr6\_massive}\ is identical except that Sgr is $1.5$ times more massive. \texttt{pmr6\_faster}\ has the fiducial Sgr mass but a higher initial infall speed. \texttt{pmr6\_host}\ is an isolated-host control, which isolates the disc evolution that would have occurred anyway.

\section{Orbit and mass loss of the Sagittarius-like dwarf galaxy}
\label{sec:orbit_mass_loss}
Here, we determine the relative orbit of the Mw-Sgr systems in Section~\ref{sec:orbit}, and how much mass Sgr loses as the orbit evolves in Section~\ref{sec:mass_loss}. For time, we work in terms of stellar age, which ranges from $0$ (the present day) to $10$~Gyr (the age of the oldest surviving star formed in the simulation).

\subsection{Identifying the two galaxies and the orbit}
\label{sec:orbit}

We load and analyse snapshots using \texttt{pynbody} \citep{pontzen+13}, and separately identify star, gas, and dark matter particles born in the MW system from those born in the Sgr system using their particle IDs, which are assigned according to a fixed numbering convention at the start of each simulation. This tag records only where a particle was born: at this stage, the Sgr-tagged set includes all satellite-origin material, whether or not it remains gravitationally bound to the remnant later on.

\begin{figure}
    \centering
    \includegraphics[width=\linewidth]{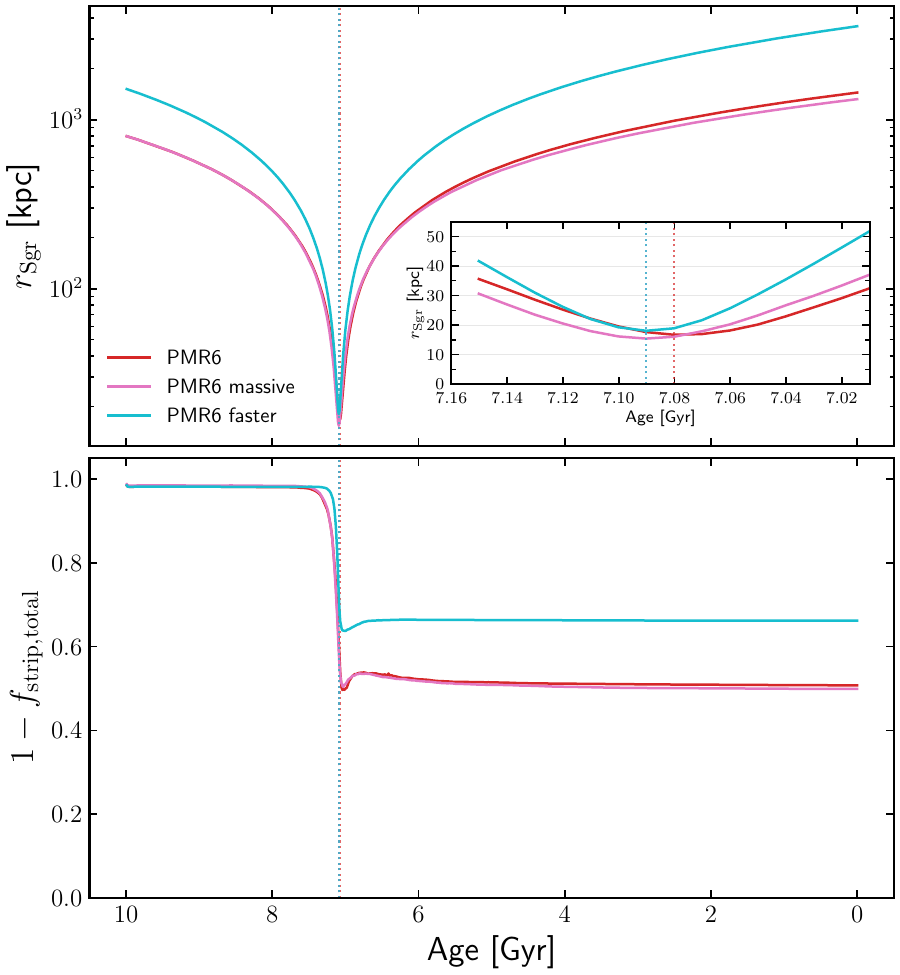}
    \caption{Orbital and mass-loss histories of Sgr in the three simulations that contain a satellite, as a function of age, i.e. time before the present day. Age decreases to the right, so time runs from left to right. \emph{Top:} the Sgr--MW separation $r_{\rm Sgr}$, on a logarithmic scale. The inset re-plots the passage on a linear scale over ages $7.16$--$7.01$~Gyr.
    \emph{Bottom:} bound fraction of all Sgr-tagged material (stars, gas, and dark matter combined), $1-f_{\rm strip,total}$, with $f_{\rm strip,total}$ as defined in Equation~\eqref{eq:fstrip}. The curves begin at $\simeq0.98$ rather than unity because a small fraction of the tagged material is unbound from the outset. Dotted vertical lines mark each run's pericentre, at an age of $\simeq7.1$~Gyr; all three lie within $11$~Myr of one another and so overlap. In every case the satellite makes a single unbound passage, and essentially all of the mass loss occurs during it. The more massive satellite loses the same fraction of its mass as the fiducial one, whereas the faster satellite, which spends less time near pericentre, loses substantially less.}
    \label{fig:sgr_orbit_massloss}
\end{figure}

We then centre each snapshot and rotate it face-on using \texttt{pynbody}'s \texttt{faceon} routine, applied to all pre-existing and newly-formed stars belonging to the MW thin and thick discs. The spatial centre used for this alignment is computed using the standard iterative shrinking-sphere algorithm \citep{power+03}. The velocity zero-point, however, is defined as the mass-weighted mean velocity of the entire disc-star population used for alignment, rather than of the small inner subset of particles used by the shrinking-sphere algorithm's default velocity centring; this avoids the alignment frame being biased by the disc's inner bar/bulge kinematics rather than its true bulk motion.

At each snapshot, we estimate the centre of Sgr, using the same iterative shrinking-sphere method applied to all Sgr-tagged star particles, irrespective of whether they are still bound. Since the MW system is by construction centred at $(0,0,0)$~kpc after the alignment step above, the separation between the two centres directly gives the Sgr--MW orbital separation, $r_{\rm Sgr}$, at that snapshot.

Figure~\ref{fig:before_during_after} shows the host disc in projection around the pericentric passage. The top plot of Figure~\ref{fig:sgr_orbit_massloss} shows the resulting orbital histories. In all three simulations, the satellite begins far outside the host and makes a single pericentric passage at an age of $7.1$~Gyr ($\tau_{\rm peri}$, Table~\ref{tab:orbit_massloss}). The infall is close to polar: averaged over the approach, the satellite centre lies only $4.5^{\circ}$ from the disc's rotation axis in \texttt{pmr6}\ and \texttt{pmr6\_massive}, and $1.8^{\circ}$ in \texttt{pmr6\_faster}\ ($\theta_{\rm polar}$, Table~\ref{tab:orbit_massloss}). The pericentric distances are $16.7$, $15.4$ and $18.1$~kpc for \texttt{pmr6}, \texttt{pmr6\_massive}\ and \texttt{pmr6\_faster}\ respectively, and the corresponding speeds at pericentre are $466$, $475$ and $634$~km~s$^{-1}$ ($r_{\rm peri}$ and $v_{\rm peri}$, Table~\ref{tab:orbit_massloss}). These values are robust to the $10$-Myr snapshot cadence (a parabolic refinement through the minimum changes $r_{\rm peri}$ by $\leq0.2$~kpc and $v_{\rm peri}$ by $\leq1$~km~s$^{-1}$) and to the choice of satellite centre (star-based, dark-matter-based and inner-COM centres agree to $\leq0.1$~kpc at pericentre).

These speeds make the encounter impulsive: the satellites cross the disc at $2.4$--$3.3$ times the local circular speed, $v_{\rm c}\simeq
190$~km~s$^{-1}$, which we measure from the total mass enclosed within a sphere of the pericentric radius in the pre-passage disc, and at $\sim50$ times the vertical velocity dispersion of the disc stars they perturb. More precisely, the encounter is impulsive where the force is delivered quickly compared with the stars' vertical oscillations \citep{spitzer58,binney_tremaine08}. The satellite delivers its kick over $t_{\rm enc}\sim 2r_{\rm peri}/v_{\rm peri}\simeq60$--$70$~Myr, while the vertical oscillation period of the pre-passage disc, $T_z = 2\pi/\sqrt{4\pi G\rho_0(R)}$ with $\rho_0$ the midplane density, rises from $\simeq100$~Myr at $R=9$~kpc to $\simeq190$~Myr at $R=15$~kpc and $\gtrsim250$~Myr beyond. Stars at and outside the crossing radius therefore receive their kick in a small fraction of a vertical period, whereas inside $R\sim7$~kpc, where $T_z\lesssim t_{\rm enc}$, the response is adiabatic and the inner disc is partially shielded from the encounter.

After pericentre, the separation increases monotonically to the end of the simulation, and the satellite does not return. The interaction modelled here is thus a single flyby, and the disc structure analysed in Section~\ref{sec:sat_impacts} is the fossil record of one passage rather than the cumulative effect of repeated encounters.

\begin{table}
    \centering
    \caption{Orbital and mass-loss properties of Sgr in the three simulations containing a satellite. $\tau_{\rm peri}$ is the age (time before the present day) of the pericentric passage; $r_{\rm peri}$ is the minimum of the three-dimensional MW-Sgr separation $r_{\rm Sgr}$; $v_{\rm peri}$ is the satellite's speed at that moment; $\theta_{\rm polar}$ is the mean angle of the satellite centre from the disc's rotation axis over the approach; $M_{\rm bound,0}$ and $M_{\rm bound,f}$ are the bound masses at the first and last snapshot; and $\Delta t_{\rm strip}$ is the duration of the mass-loss episode. The stripped fraction is $f_{\rm strip,total}$ at the final snapshot, Equation~\eqref{eq:fstrip}.}
    \label{tab:orbit_massloss}
    \begin{tabular}{lccc}
        \hline
        & \texttt{pmr6} & \texttt{pmr6\_massive} & \texttt{pmr6\_faster} \\
        \hline
        $\tau_{\rm peri}$ [Gyr]                  & 7.1 & 7.1  & 7.1 \\
        $r_{\rm peri}$ [kpc]                     & 16.7 & 15.4  & 18.1 \\
        $v_{\rm peri}$ [km~s$^{-1}$]            & 466  & 475   & 634  \\
        $\theta_{\rm polar}$ [deg]              & 4.5  & 4.5   & 1.8  \\
        $M_{\rm bound,0}$ [$10^{10}$\,M$_\odot$] & 7.08 & 10.60 & 7.08 \\
        $M_{\rm bound,f}$ [$10^{10}$\,M$_\odot$] & 3.65 & 5.35  & 4.76 \\
        $f_{\rm strip,total}$                    & 0.493 & 0.501 & 0.338 \\
        $\Delta t_{\rm strip}$ [Gyr]             & 0.2 & 0.2  & 0.1 \\
        \hline
    \end{tabular}
\end{table}

\subsection{Bound mass and mass loss}
\label{sec:mass_loss}

Determining how much mass the satellite loses due to tidal stripping requires going beyond the birth tags of Section~\ref{sec:orbit} and asking which of the Sgr-tagged material is still gravitationally bound to the remnant. We first select candidate Sgr stars, gas, and dark matter within 50~kpc of the shrinking-sphere system centre. From this candidate set, we construct a multipole expansion of the Sgr gravitational potential using \texttt{Agama} \citep{vasiliev19}, expanded to order $l_{\rm max}=8$ with no assumed symmetry, in units of kpc, km~s$^{-1}$, and M$_\odot$.

For each particle, we then calculate the specific energy as
\begin{equation}
E = \frac{1}{2} v_{\rm rel}^{2} + \Phi_{\rm Sgr},
\end{equation}
where $v_{\rm rel}$ is the particle's velocity relative to the mass-weighted bulk velocity of the current candidate bound set, and $\Phi_{\rm Sgr}$ is the remnant potential evaluated at the particle's position. Particles with $E<0$ are classified as bound.

We employ an iterative unbinding procedure. Following the initial energy calculation, we remove particles classified as unbound and recompute both the potential and the bulk velocity using only the remaining bound material, then repeat the energy calculation and classification with this updated potential and bulk velocity. The iteration continues until the total bound mass changes by less than a fractional tolerance of $10^{-3}$ between successive iterations, or a maximum of 20 iterations is reached.

We calculate the bound mass of component $X$ (stars, gas, dark matter, or their sum) at time $t$ as
\begin{equation}
M_{\rm bound,X}(t) = \sum_{i\in X,\,E_i<0} m_i,
\end{equation}
and the corresponding stripped mass fraction as
\begin{equation}
f_{\rm strip,X}(t) = \frac{M_{\rm total,X}(t) - M_{\rm bound,X}(t)}{M_{\rm total,X}(t)},
\label{eq:fstrip}
\end{equation}
where $M_{\rm total,X}(t)$ is the total mass of all Sgr-tagged particles of component $X$ present at time $t$ (bound and stripped combined). This tagged set is conserved to better than one part in $10^{5}$ over the full run, so the denominator of $f_{\rm strip}$ is stable; the numerator is insensitive to the $50$~kpc candidate aperture because the converged bound remnant is always far more compact than the aperture. $f_{\rm strip,total}$ then denotes the stripped mass fraction of all the components together. Note that $f_{\rm strip,total}$ is already $0.012$--$0.015$ in the first snapshot. This reflects the smoothly truncated outskirts of the realized satellite halo (Table~\ref{tab:sim_params}): the initially unbound material lies entirely beyond $50$~kpc from the remnant and never belonged to the bound body.

We evaluate $f_{\rm strip}$ at fixed $t$ rather than relative to the initial bound mass, because the two measures differ substantially for the individual components. The satellite continues to form stars from its own gas throughout the run, so its bound stellar mass more than doubles; a measure normalized to the initial bound mass would therefore return a negative stripped fraction for the stellar component.

The bottom panel of Figure~\ref{fig:sgr_orbit_massloss} shows the resulting mass-loss histories, summarized in Table~\ref{tab:orbit_massloss}. The bound fraction is constant to within a few per cent until shortly before pericentre, falls sharply over a single episode lasting $0.2$, $0.2$ and $0.1$~Gyr in \texttt{pmr6}, \texttt{pmr6\_massive}\ and \texttt{pmr6\_faster}\ respectively ($\Delta t_{\rm strip}$, Table~\ref{tab:orbit_massloss}), and is then flat for the remaining $\sim7$~Gyr. Stripping is therefore confined almost entirely to the passage itself: the bound mass at the end of the simulation is set within a few hundred Myr of $\tau_{\rm peri}$, and the remnant neither regains nor continues to lose appreciable mass thereafter.

Comparing the runs isolates the roles of satellite mass and encounter speed. Stripped fractions quoted here and in Table~\ref{tab:orbit_massloss} are $f_{\rm strip,total}$ at the final snapshot, as defined in Equation~\eqref{eq:fstrip}. \texttt{pmr6}\ and \texttt{pmr6\_massive}\ follow nearly identical trajectories, cross at nearly the same speed, and lose nearly identical fractions of their mass ($49.3$ and $50.1$~per cent) despite the latter being $50$~per cent more massive, indicating that the stripped fraction is governed by the orbit rather than by the satellite mass. By contrast \texttt{pmr6\_faster}, which crosses at a slightly larger pericentric distance and at a substantially higher speed ($634$ compared with $466$~km~s$^{-1}$, Table~\ref{tab:orbit_massloss}), retains far more of its mass and loses only $33.8$~per cent. Its stripping episode is also the shortest, at $0.1$~Gyr, so the reduced mass loss follows naturally from the shorter time spent near pericentre.

\section{Fossil impacts on the phase-space structure}\label{sec:sat_impacts}

In this Section, we discuss the present-day density and kinematic structure of the disc of the MW systems in all four simulations. Throughout, we analyse the newly-formed thin-disc stars in the present-day snapshot only. Stellar positions and velocities are expressed throughout in cylindrical coordinates, $(R,\phi,z)$ and $(v_R,v_\phi,v_z)$ respectively. All panels below restrict to $R>6$~kpc, except Figure~\ref{fig:radial_density_profiles}, which spans the full range $R=0$--$30$~kpc to show the central density excess. This excess is the nuclear stellar disc, which is too strong, but irrelevant for our science case.

\subsection{Present-day density structure of the thin disc}

\begin{figure*}
    \centering
    \includegraphics[width=\linewidth]{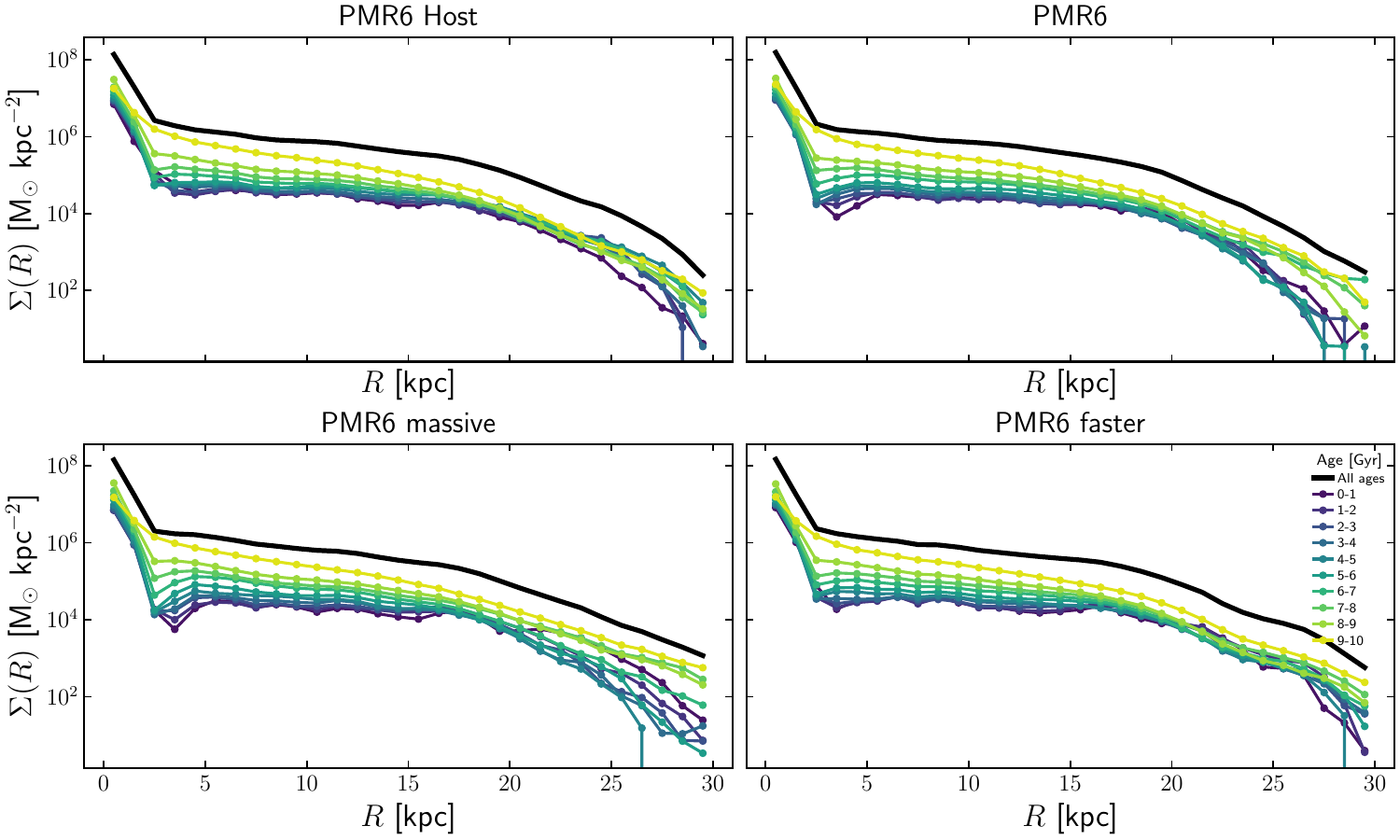}
    \caption{
    Present-day radial surface mass density profiles of newly-formed thin-disc stars in the four simulations. The black curve shows the total surface mass density, while coloured curves show stars in different age intervals. At fixed radius, older age bins generally sit above younger ones, reflecting a declining star formation history with time.
    }
    \label{fig:radial_density_profiles}
\end{figure*}

The present-day radial surface mass density profiles shown in Figure~\ref{fig:radial_density_profiles} are computed both in stellar age bins and for the full combined stellar population. Stars are binned into annuli of galactocentric radius $R$, $1$~kpc wide and spanning $R=0$ to $30$~kpc, over all heights $z$ above and below the plane. The stellar age bins are $1$~Gyr wide, and span $0$ to $10$~Gyr. The surface mass density in the $j$-th annulus $[R_j,R_{j+1})$ is
\begin{equation}
    \Sigma(R_j) = \frac{1}{\pi\left(R_{j+1}^{2}-R_{j}^{2}\right)}\sum_{i\,:\,R_j \le R_i < R_{j+1}} m_i ,
    \label{eq:SigmaR}
\end{equation}
where the sum runs over the masses $m_i$ of stars falling within that annulus, and the normalization is the annulus area. The same binning is applied independently within each stellar age interval to produce the coloured curves in Figure~\ref{fig:radial_density_profiles}, and to the full population to produce the black curve. 

The broad shape is similar for each simulation. There is a steep central excess within the inner few kpc, a localized dip near $R\sim3$~kpc, a roughly flat region out to $R\sim17$~kpc, followed by a steep decline beyond it. Because this shape is present in the isolated host as well as the three satellite runs, it is not a response to the Sgr passage. Since only newly-formed thin-disc stars are shown here, it instead reflects where these stars form within the gas disc over the course of the simulation. At fixed radius, older age bins sit systematically above younger ones, reflecting the disc's declining star-formation history as its gas reservoir is depleted over the run. In \texttt{pmr6} and \texttt{pmr6\_massive}, there appears to be a split in the profiles around the pericentric passage beyond 25 kpc, however in the latter, it manifests more as a general broadening of the profiles in the very outer disc.

To quantify how the disc's radial distribution evolves with stellar age, we define a mass-enclosed radius $R_{50}$, computed independently for each stellar age bin and simulation. We use this instead of an exponential scale radius as the surface-density profiles are not well described by a single exponential. Stars in a given subset are ranked by increasing galactocentric radius $R$, and the enclosed mass within radius $R$ is written $M(<R) = \sum_{i\,:\,R_i<R} m_i$, where the sum runs over the stars in the subset and $m_i$ is the mass of the $i$-th star. Writing $M_{\rm tot} = \sum_i m_i$ for the subset's total mass, $R_{50}$ is defined as the radius enclosing half of it, i.e.
\begin{equation}
     M(<R_{50}) = \frac{1}{2} M_{\rm tot}.
     \label{eq:R50}
\end{equation}
This is solved numerically by linear interpolation of the cumulative mass profile $M(<R)$.

\begin{figure}
    \centering
    \includegraphics[width=\linewidth]{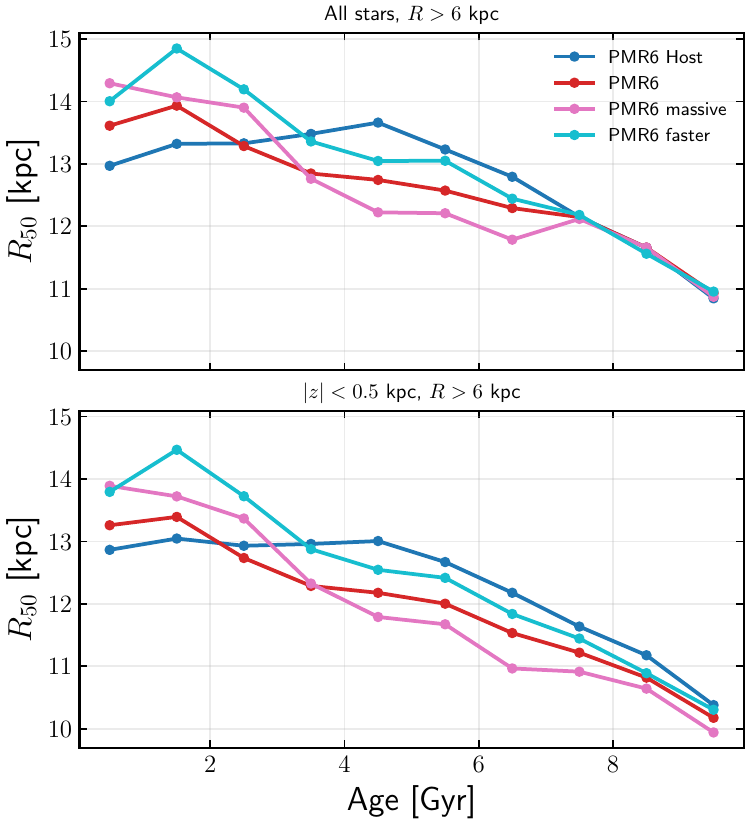}
    \caption{
    Mass-enclosed radius $R_{50}$ of newly-formed thin-disc stars as a function of stellar age, for all such stars at $R>6$~kpc (top) and for those close to the plane, $|z|<0.5$~kpc (bottom). $R_{50}$ generally increases from $\sim11$~kpc for the oldest stars to $\sim14$--$15$~kpc for the youngest stars in all four simulations, consistent with inside-out disc growth. The host however shows a slight decrease after 5 Gyr. \texttt{pmr6} and \texttt{pmr6\_faster} show a change in the rate at which $R_{50}$ increases with time at $7.1$ Gyr, but only \texttt{pmr6\_massive} shows a dip.
    }
    \label{fig:scale_radius_vs_age}
\end{figure}

Figure~\ref{fig:scale_radius_vs_age} shows $R_{50}$ as a function of stellar age, calculated for all stars at $R>6$~kpc (top) and for stars restricted to the plane, $|z|<0.5$~kpc (bottom), in all four simulations. In both panels, $R_{50}$ declines with age in a broadly similar way across all four runs, from $\sim14$--$15$~kpc for the youngest stars (age $<1$~Gyr) to $\sim10$--$11$~kpc for the oldest (age $9$--$10$~Gyr). This is a signature of inside-out disc growth.

Superposed on this shared trend, $R_{50}$ dips to lower values for \texttt{pmr6\_massive} compared to the other three simulations, across the $\sim3$--$7$~Gyr age range, reaching a minimum of $11.8$~kpc (top) and $11.0$~kpc (bottom) at a stellar age of $\sim6.5$~Gyr. This shows that the pericentric passage leads to a shrinking of the disc in the case of the most massive satellite (\texttt{pmr6\_massive}). In the fiducial simulation (\texttt{pmr6}) and the simulation with a faster satellite (\texttt{pmr6\_faster}), the discs continue to grow after the pericentric passage but not as much as in \texttt{pmr6\_host}, where there is no interaction with the satellite. In \texttt{pmr6\_host}, the disc stops growing beyond about five Gyr. In the case of any pericentric passage, the disc continues to grow until almost the present day, either because the gas does not get used up as quickly as in the host-only case, or because the satellite brings in fresh gas that can form stars.

The key difference between the profiles using all stars beyond $R=6$~kpc (top) and using just in-plane stars beyond $R=6$~kpc (bottom) is that before the pericentric passage, $R_{50}$ aligns exactly between all four simulations in the former, but not in the latter. 

\begin{figure*}
    \centering
    \includegraphics[width=\linewidth]{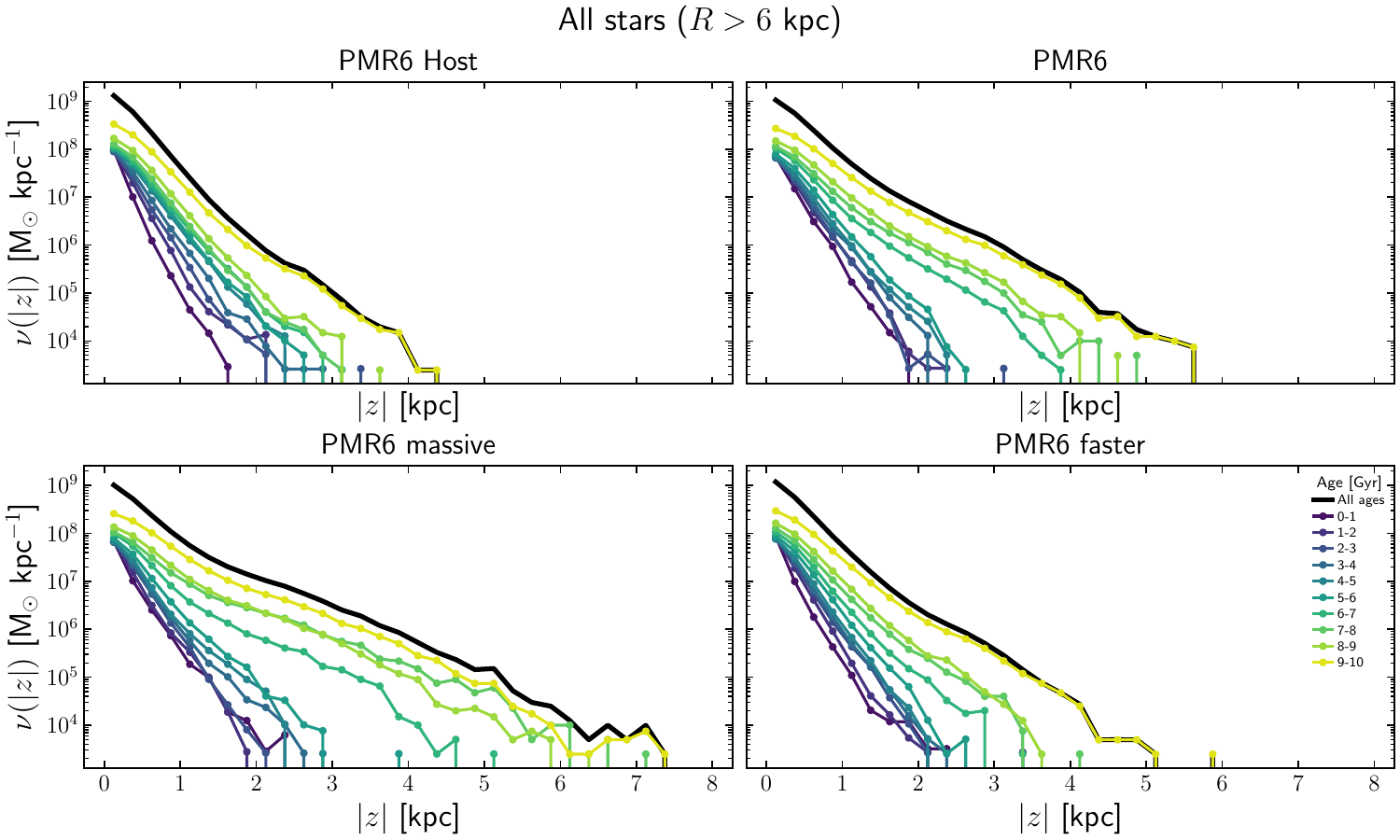}
    \caption{
    Present-day vertical mass density profiles of newly-formed thin-disc stars, restricted to $R>6$~kpc. The black curve shows all such stars, while coloured curves indicate different stellar age bins. Older populations exhibit systematically shallower, more extended profiles, reflecting progressive disc heating and the formation of a thick-disc component. Differences between simulations are most evident for the oldest populations, particularly in \texttt{pmr6\_massive}, where the oldest bins reach $|z|\sim6$--$7$~kpc compared with $\sim3$--$4$~kpc in the isolated host; Figure~\ref{fig:scale_height_vs_age} quantifies this via the mass-enclosed height $|z|_{50}$ as a function of stellar age.
    }
    \label{fig:vertical_density_profiles}
\end{figure*}

Figure~\ref{fig:vertical_density_profiles} shows the corresponding present-day vertical mass density profile, computed in the same way as Equation~\eqref{eq:SigmaR} but binning the same $R>6$~kpc stellar subset by height above the plane, $|z|$, rather than by galactocentric radius. The vertical mass density in the $k$-th height bin $[|z|_k,|z|_{k+1})$ is
\begin{equation}
    \nu(|z|_k) = \frac{1}{|z|_{k+1}-|z|_k}\sum_{i\,:\,|z|_k \le |z|_i < |z|_{k+1}} m_i ,
    \label{eq:nuz}
\end{equation}
where the sum runs over the masses of stars in that height bin. Unlike Equation~\eqref{eq:SigmaR}, this is normalized by the bin's vertical extent rather than by an area: every $|z|$ bin covers the same in-plane footprint (the full $R>6$~kpc annulus), so the cross-sectional area is a common constant that would only rescale the profile, whereas the area of a radial annulus in Equation~\eqref{eq:SigmaR} varies with $R$ and must be divided out.

In all four simulations, older stars are found at progressively larger $|z|$, a signature of disc heating. Older populations have had more time to be dynamically heated, and are therefore vertically more extended than younger ones. In \texttt{pmr6} and, more strikingly, \texttt{pmr6\_massive}, this otherwise smooth progression shows an evident jump: the age bins old enough to have existed at the $7.1$~Gyr-ago pericentric passage reach much larger $|z|$ than a smooth extrapolation of the younger bins would suggest, most dramatically in \texttt{pmr6\_massive}. In \texttt{pmr6\_faster}, the jump is much smaller.

To quantify how the disc's vertical structure evolves with stellar age, we similarly define a mass-enclosed height $|z|_{50}$, computed independently for each stellar age bin and simulation. As with $R_{50}$, this is a percentile of the mass distribution rather than a fitted scale height, so it requires no assumption that the vertical profile follows a single exponential (or any other fixed functional form). Stars in a given subset are ranked by increasing height above the plane $|z|$, and the enclosed mass within height $|z|$ is written $M(<|z|) = \sum_{i\,:\,|z|_i<|z|} m_i$, where the sum runs over the stars in the subset and $m_i$ is the mass of the $i$-th star. Writing $M_{\rm tot} = \sum_i m_i$ for the subset's total mass, $|z|_{50}$ is defined as the height enclosing half of it, i.e.,
\begin{equation}
     M(<|z|_{50}) = \frac{1}{2} M_{\rm tot} ,
     \label{eq:z50}
\end{equation}
solved numerically by linear interpolation of the cumulative mass profile $M(<|z|)$.

\begin{figure}
    \centering
    \includegraphics[width=\linewidth]{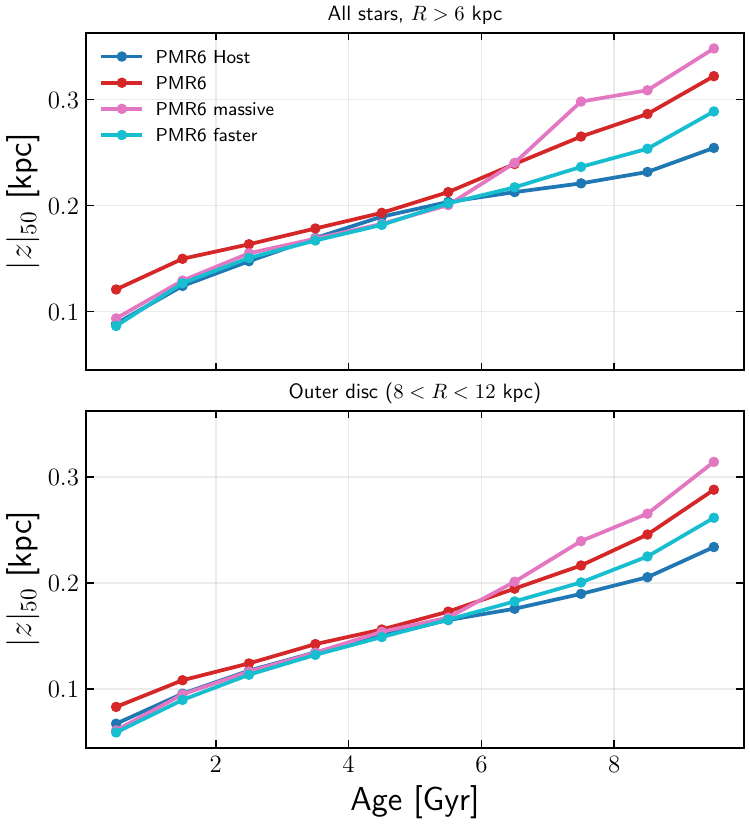}
    \caption{
    Mass-enclosed height $|z|_{50}$ of newly-formed thin-disc stars as a function of stellar age, for all such stars at $R>6$~kpc (top) and for the outer disc, $8<R<12$~kpc (bottom). All four simulations, including the host, show $|z|_{50}$ increasing with age, the signature of ongoing secular disc heating. The three Sagittarius runs track the host closely for stellar ages below $\sim5$--$6$~Gyr, then depart upward, with \texttt{pmr6\_massive} diverging first and most strongly. By the oldest age bin, $|z|_{50}$ reaches $0.35$~kpc in \texttt{pmr6\_massive} against $0.26$~kpc in the host (all stars, top panel) -- a $\sim37$~per cent excess. The onset of this divergence coincides with the $7.1$~Gyr stellar age of the passage: stars old enough to have existed at the encounter are permanently thickened relative to the host control, while younger stars are not.
    }
    \label{fig:scale_height_vs_age}
\end{figure}

Figure~\ref{fig:scale_height_vs_age} shows $|z|_{50}$ as a function of stellar age, for all stars at $R>6$~kpc (top) and for the outer disc, $8<R<12$~kpc (bottom). Both panels track each other closely in shape: $|z|_{50}$ rises steadily with age in all four simulations from $\sim0.06$--$0.09$~kpc for the youngest stars to a shared value of $\sim0.20$~kpc at a stellar age of $\sim5.5$~Gyr, with the four runs essentially indistinguishable up to this point.

Beyond a stellar age of $\sim6$--$7$~Gyr, the runs separate cleanly and in the same order in both panels: \texttt{pmr6\_massive} rises most steeply, \texttt{pmr6} next, then \texttt{pmr6\_faster}, with the isolated host rising most gradually throughout. The steepest part of \texttt{pmr6\_massive}'s divergence -- from $|z|_{50}\simeq0.22$ to $0.30$~kpc (top panel) -- occurs between the age bins immediately straddling the $7.1$~Gyr passage age, coincident with the jump already seen directly in Figure~\ref{fig:vertical_density_profiles}. By the oldest bin, the ordering is preserved in both panels. $|z|_{50}$ reaches $0.35$, $0.32$ and $0.29$~kpc in \texttt{pmr6\_massive}, \texttt{pmr6}\ and \texttt{pmr6\_faster}\ (all stars, top panel), against $0.26$~kpc in the host, and $0.31$, $0.29$ and $0.26$~kpc respectively against $0.23$~kpc in the host in the outer-disc bottom panel. That the same ordering and the same $\sim7$~Gyr onset appear in both an all-heights and an outer-disc-only selection indicates this is a genuine property of the disc's vertical structure, not an artefact of the particular radial window chosen.

\subsection{Present-day velocity dispersion structure of the thin disc}

Using the vertical velocity component $v_z$ defined above, we compute radial profiles of the vertical stellar velocity dispersion, $\sigma_z$. We focus on the vertical component because it carries the cleanest fossil. The radial velocity dispersion of pre-passage stars also shows an excess over the control ($\Delta\sigma_R\simeq+4$~km~s$^{-1}$ in all three satellite runs), but with $\sim3\times$ larger azimuthal variance from spiral streaming and no ordering with satellite mass or speed, indicating that it is dominated by the satellite-induced spiral response rather than by the direct kick; $\sigma_\phi$ shows only marginal excesses ($\lesssim1.5$~km~s$^{-1}$).

We divide the disc into radial bins spanning \(6 < R < 30\,\mathrm{kpc}\), the same radial range used throughout this Section. Unlike the fixed-width annuli used for the radial surface mass density profile (Equation~\eqref{eq:SigmaR}), here we choose the bins using an equal-number scheme: $12$ bins defined by the quantiles of the radial distribution, so that each contains approximately the same number of stars. This avoids the outer bins being left with too few stars for a reliable dispersion estimate, as would happen with the fixed-width scheme given the steeply declining density of stars at large $R$. Within each radial bin, we further subdivide the stars into age intervals.

Prior to estimating the velocity dispersion, we apply a \(5\sigma\) clip to remove extreme outliers. We then determine the intrinsic velocity dispersion using a maximum-likelihood estimator that accounts for measurement uncertainties, following the method of \citet{sanders+18} and \citet{das+24}. Assuming a Gaussian vertical velocity distribution with mean \(\mu_z\) and intrinsic dispersion \(\sigma_z\), we maximize the likelihood by solving
\begin{equation}
\sum_i
\left[
\frac{(v_{z,i}-\mu_z)^2}
     {(\sigma_z^2+\epsilon_i^2)^2}
-
\frac{1}
     {\sigma_z^2+\epsilon_i^2}
\right]
=0,
\end{equation}
where \(v_{z,i}\) is the vertical velocity of the \(i\)-th star and \(\epsilon_i\) is its associated velocity uncertainty. The simulated velocities carry no measurement error, so we assign each star a nominal placeholder uncertainty of \(\epsilon_i = 0.05\,|v_{z,i}|\); since \(\epsilon_i\ll\sigma_z\), its effect on the recovered dispersions is below $0.2$~per cent. We obtain the solution numerically using Brent's root-finding algorithm, and apply this procedure independently within each age--radius bin, producing a profile \(\sigma_z(R)\) for each age interval.

We estimate the uncertainty on the recovered dispersion as
\begin{equation}
\delta \sigma_z \simeq
\frac{\sigma_z}{\sqrt{2N}},
\end{equation}
where \(N\) is the number of stars in the corresponding age--radius bin. We only report dispersion measurements for bins containing at least 30 stars. This formal uncertainty is negligible at the bin sizes used here; uncertainties quoted on dispersion differences between runs are instead derived from the scatter of $\sigma_z$ across eight azimuthal sectors, which captures the variance contributed by spiral and other non-axisymmetric structure.

\begin{figure*}
    \centering
    \includegraphics[width=\linewidth]{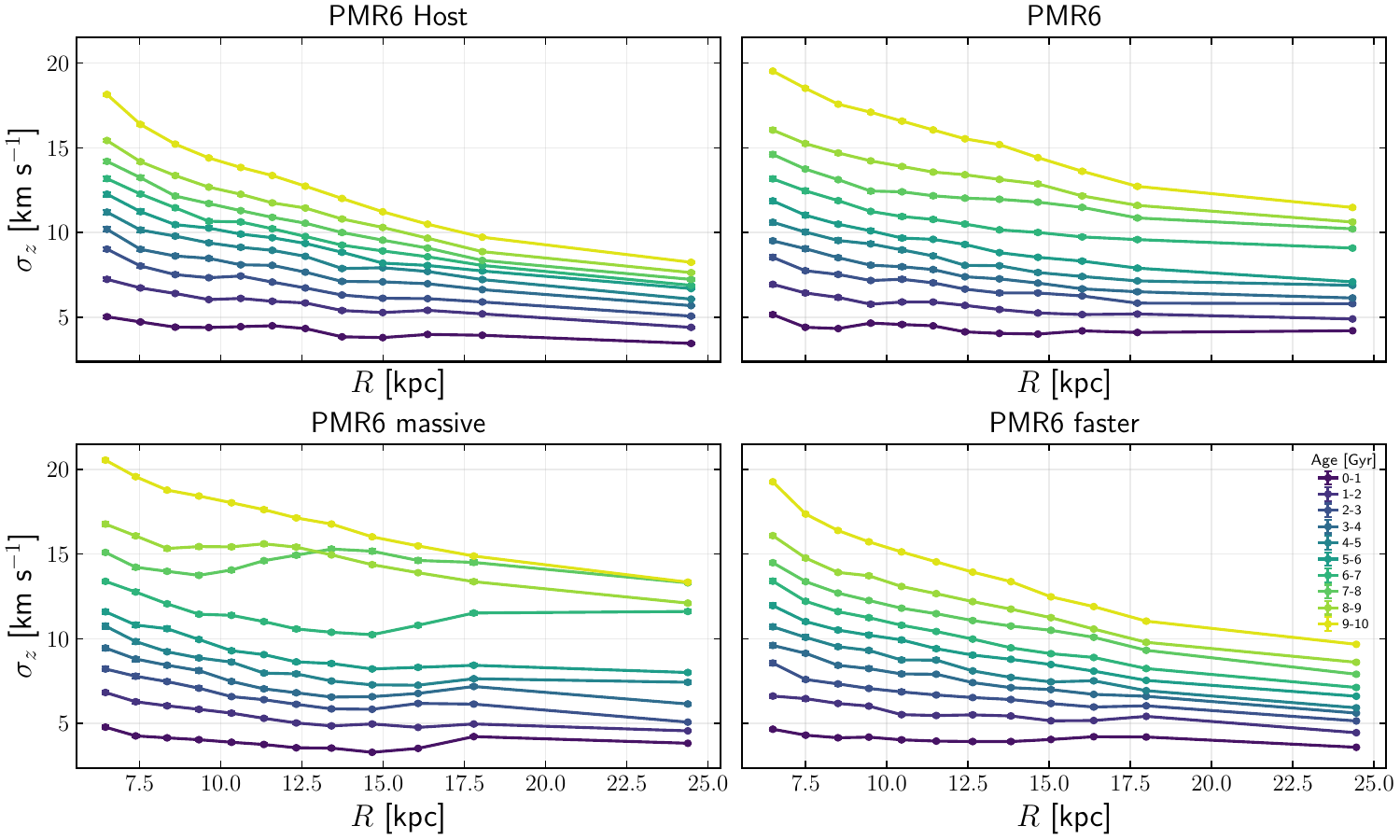}
    \caption{
    Radial profiles of the vertical velocity dispersion $\sigma_z$ of newly-formed thin-disc stars, one panel per simulation. Coloured curves show $1$~Gyr-wide stellar age bins; the dispersions are measured in twelve equal-number radial bins spanning $6<R<30$~kpc, error bars show $\delta\sigma_z=\sigma_z/\sqrt{2N}$, and only bins containing at least $30$ stars are plotted. At every radius, older populations have systematically larger $\sigma_z$, reflecting progressive disc heating. In the isolated host and \texttt{pmr6\_faster}\ every age bin declines smoothly and monotonically with radius, while in \texttt{pmr6}\ the oldest bins level off at $\sim11$--$12$~km~s$^{-1}$ in the outer disc. \texttt{pmr6\_massive}\ departs qualitatively from this pattern: the age bins old enough to have existed at the $7.1$~Gyr-ago passage show a localized, non-monotonic bump near $R\sim12$--$15$~kpc, close to that run's pericentric radius of $15.4$~kpc, indicating that the excess heating is deposited preferentially near the satellite's disc-crossing radius, as expected from the adiabatic shielding of the inner disc (Section~\ref{sec:orbit}).
    }
    \label{fig:sigmaz_profiles}
\end{figure*}

In the host and in \texttt{pmr6\_faster}, every age bin in Figure~\ref{fig:sigmaz_profiles} declines smoothly and monotonically with radius, with no irregular features at any $R$; the oldest bin falls from $\sim18$--$20$~km~s$^{-1}$ at $R=6$~kpc to $\sim8$--$10$~km~s$^{-1}$ at $R=25$~kpc in both runs. \texttt{pmr6}\ is similarly smooth, though its oldest bins fall off less steeply than the host's, levelling off at $\sim11$--$12$~km~s$^{-1}$ in the outer disc rather than continuing to decline. \texttt{pmr6\_massive}\ departs qualitatively from this pattern: the age bins old enough to have existed at the passage show a localized, non-monotonic bump rather than a smooth decline. The $7$--$8$~Gyr bin falls from $\sim16.8$~km~s$^{-1}$ at $R=6$~kpc to a local minimum of $\sim13.7$~km~s$^{-1}$ near $R\sim9$--$10$~kpc, then rises back to a local maximum of $\sim15.3$~km~s$^{-1}$ at $R\sim12$--$14$~kpc before resuming its decline; the $6$--$7$~Gyr bin shows the same pattern shifted outward, bottoming out at $\sim10.3$~km~s$^{-1}$ near $R\sim14$--$15$~kpc before rising to a plateau of $\sim11.6$~km~s$^{-1}$ for $R\gtrsim18$~kpc. Both bumps sit close to this run's pericentric distance of $15.4$~kpc (Table~\ref{tab:orbit_massloss}), suggesting the excess heating is deposited preferentially near the satellite's disc-crossing radius rather than uniformly across the disc, as expected from the adiabatic shielding of the inner disc estimated in Section~\ref{sec:orbit}.

\begin{figure}
    \centering
    \includegraphics[width=\linewidth]{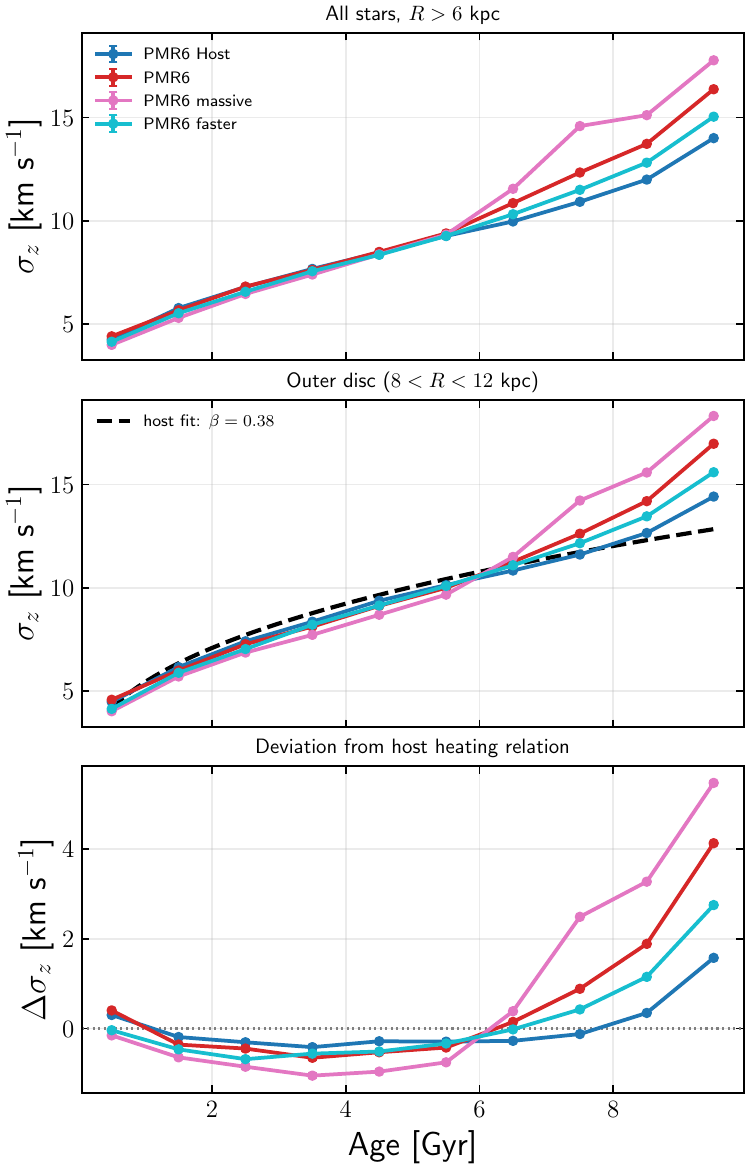}
    \caption{
    Vertical velocity dispersion $\sigma_z$ of newly-formed thin-disc stars as a function of stellar age, over all heights. \emph{Top:} all stars at $R>6$~kpc. \emph{Middle:} the outer disc, $8<R<12$~kpc, with a power-law heating relation $\sigma_z=A(\tau/\tau_0)^{\beta}$ fitted to the isolated host only ($\beta=0.38$), following the quiescent-heating picture of \citet{sanders+15}. \emph{Bottom:} the residual $\Delta\sigma_z(\tau)=\sigma_{z,\mathrm{sim}}(\tau)-\sigma_{z,\mathrm{host\ fit}}(\tau)$ for all four simulations, including the host's own residual against its fit. In all three panels, the three satellite runs track the host closely below a stellar age of $\sim6$~Gyr and depart above it, led by \texttt{pmr6\_massive}. In the oldest age bin (top panel), $\sigma_z$ reaches $17.8$, $16.4$ and $15.0$~km~s$^{-1}$ in \texttt{pmr6\_massive}, \texttt{pmr6}\ and \texttt{pmr6\_faster}\ respectively, against $14.5$~km~s$^{-1}$ in the host; in the bottom panel, the same ordering gives $\Delta\sigma_z\simeq+5.4$, $+4.1$ and $+2.8$~km~s$^{-1}$ against $+1.6$~km~s$^{-1}$ for the host's own residual against its fit.
    }
    \label{fig:sigma_z_vs_age_heating}
\end{figure}

Figure~\ref{fig:sigma_z_vs_age_heating} shows $\sigma_z$ as a function of stellar age in three complementary ways. The top panel, for all stars at $R>6$~kpc, and the middle panel, restricted to the outer disc ($8<R<12$~kpc), both show the same qualitative behaviour already seen for $|z|_{50}$ (Figure~\ref{fig:scale_height_vs_age}): $\sigma_z$ rises steadily with age up to a stellar age of $\sim6$~Gyr, with all four simulations indistinguishable over this range, then separates cleanly beyond it, with \texttt{pmr6\_massive}\ departing first and most strongly, followed by \texttt{pmr6}\ and \texttt{pmr6\_faster}, and the host rising most gradually throughout. By the oldest age bin, $\sigma_z$ reaches $17.8$, $16.4$ and $15.0$~km~s$^{-1}$ in \texttt{pmr6\_massive}, \texttt{pmr6}\ and \texttt{pmr6\_faster}\ respectively, against $14.5$~km~s$^{-1}$ in the host (top panel) -- a $\sim23$~per cent excess for the most massive satellite.

The middle panel overlays a power-law heating relation, $\sigma_z(\tau) = A(\tau/\tau_0)^{\beta}$, fitted to the host run alone in the outer-disc window, following the quiescent-heating picture of \citet{sanders+15}; this gives $\beta=0.38$, broadly consistent with secular disc heating in the absence of a perturber. The bottom panel shows the residual of each run against this host fit, $\Delta\sigma_z(\tau)$. For stellar ages below $\sim6$~Gyr, all three satellite runs scatter around the host fit at the $\lesssim1$~km~s$^{-1}$ level, comparable to the host's own residual against its fit -- i.e. consistent with no excess heating at these ages. Beyond $\sim6$--$7$~Gyr, all three depart upward, most strongly in \texttt{pmr6\_massive}, which reaches $\Delta\sigma_z\simeq+5.4$~km~s$^{-1}$ in the oldest bin, against $+4.1$ and $+2.8$~km~s$^{-1}$ in \texttt{pmr6}\ and \texttt{pmr6\_faster}, compared with $+1.6$~km~s$^{-1}$ for the host's own residual against its fit. This excess-heating signature, confined to stars old enough to have existed at the $7.1$~Gyr-ago passage and absent for younger, post-passage populations, mirrors the fossil signature already identified via $|z|_{50}$ and the vertical density profiles, and confirms it is not an artefact of the particular radial window used for those metrics.

Figures~\ref{fig:scale_height_vs_age}--\ref{fig:sigma_z_vs_age_heating} show that stars old enough to have been present at the $7.1$~Gyr-ago passage are permanently thicker and dynamically hotter in the Sagittarius runs than in the isolated host, while younger, post-passage stars are not, in all three satellite simulations. \texttt{pmr6\_massive} shows the largest departures throughout, and \texttt{pmr6\_faster} shows the smallest.

\section{Discussion}
\label{sec:discussion}

Here, we first show that the strength of the fossil heating across our three satellite runs follows the scaling predicted by the impulse approximation. We then compare the impacts on the phase-space structure found in our simulations with other simulations of satellite-driven disc heating, and then with observational evidence for an early encounter between the Milky Way and Sagittarius. Then we assess whether Sagittarius could plausibly have had the orbit and mass such an encounter would require. Finally, we discuss the limitations of our work.

\subsection{Scaling of the fossil heating with satellite mass and encounter speed}
\label{ssec:heating_scaling}

The three satellite runs let us test directly whether the fossil signature scales as the impulse approximation predicts. For stars old enough to predate the passage, we measure the heating as the excess over the isolated host, averaged over the age bins spanning $7$--$10$~Gyr: $\Delta\sigma_z = 1.8$, $3.5$ and $0.8$~km~s$^{-1}$ (each $\pm0.1$, from the azimuthal-sector scatter of Section~\ref{sec:sat_impacts}, with a $\lesssim0.3$~km~s$^{-1}$ run-to-run systematic measured from the post-passage stars; Section~\ref{ssec:caveats}), and $\Delta|z|_{50} = 0.06$, $0.08$ and $0.02$~kpc, for \texttt{pmr6}, \texttt{pmr6\_massive}\ and \texttt{pmr6\_faster}\ respectively ($R>6$~kpc; the outer-disc window $8<R<12$~kpc gives the same ratios, and all values are unchanged to $0.2$~km~s$^{-1}$ if remeasured $1$~Gyr before the final snapshot). In the impulsive limit the velocity kick scales as $\Delta v \propto GM/(b\,v)$ \citep{chew_wick52,binney_tremaine08}, so the vertical energy deposited should scale as $\Delta\sigma_z^{2} \propto (GM/v)^{2}$ at fixed encounter geometry. This expression strictly describes a point-mass perturber, but the scalings with $M$ and $v$ carry over to our extended satellites, which share the same internal profile across the three runs. \texttt{pmr6\_massive}\ follows a nearly identical orbit to \texttt{pmr6}\ with $1.5$ times the mass, so its energy gain should be larger by $(1.50\times466/475)^{2} = 2.2$; we measure $2.05\pm0.15$. \texttt{pmr6\_faster}\ carries the fiducial infall mass but crosses $36$~per cent faster, predicting a ratio of $0.54$ relative to \texttt{pmr6}; we measure $0.45\pm0.07$. The measured heating therefore follows the impulsive kick scaling to better than $\sim20$~per cent, as close as can reasonably be expected, given that the satellites lose a large fraction of their mass during the passage itself. The agreement is best using the infall mass rather than the pericentric bound mass, as expected physically. Material stripped mid-passage is still close to the remnant when the kick is delivered.

This scaling implies that a measured $\Delta\sigma_z$ constrains only the combination $M/v_{\rm peri}$, not the satellite mass and crossing speed separately -- a degeneracy inherent in any impulse-based inference from the fossil record, such as that of \citet{das+24}, which independent constraints on the orbit must break. In reality the degeneracy extends to the pericentric radius $b$ as well, but the orbit, and hence $b$, is usually far better constrained than the perturber's mass, so the degeneracy of practical importance is that between $M$ and $v_{\rm peri}$.

All of this rests on the encounter being impulsive -- true by construction (Section~\ref{sec:simulation_setup}) and verified for the disc outside the crossing radius by the timescale comparison of Section~\ref{sec:orbit}. Within this regime, the calibration can be turned around and applied to the real Milky Way. \citet{das+24} measure a jump of $\Delta\sigma_z\simeq4$~km~s$^{-1}$ for stars beyond $\sim11$~kpc -- a departure of the old stars from the smooth trend traced by younger ones, with no control galaxy to subtract. The matching observable in our runs is therefore the residual above the host's fitted heating relation (Figure~\ref{fig:sigma_z_vs_age_heating}, bottom panel): $+4.1$, $+5.4$ and $+2.8$~km~s$^{-1}$ for \texttt{pmr6}, \texttt{pmr6\_massive}\ and \texttt{pmr6\_faster}\ respectively. The observed jump thus points to a perturber with $M/v_{\rm peri}\simeq1.5\times10^{8}$~M$_\odot$~km$^{-1}$~s -- a first-infall mass of $\simeq0.7\times10^{11}$~M$_\odot$ at a crossing speed of $\sim470$~km~s$^{-1}$, essentially \texttt{pmr6}. The inversion is good to a factor of $\sim1.5$ in mass: had we instead used the host-subtracted heating of pre-passage stars in the same radial window ($2.4$, $4.5$ and $0.9$~km~s$^{-1}$) -- the true deposited energy, measurable only with a control run -- the data would point to $\simeq10^{11}$~M$_\odot$, essentially \texttt{pmr6\_massive}. The difference between the two metrics is the host's own $+1.6$~km~s$^{-1}$ residual against its fit, which the observable metric cannot separate from the encounter. Both masses lie within the $0.6$--$1\times10^{11}$~M$_\odot$ range \citet{das+24} themselves infer analytically from the impulse approximation. What transfers to a disc warmer than ours is the deposited energy, not the dispersion jump itself: the cross term $\langle v_z\,\Delta v_z\rangle$ averages to zero, so the kick adds $\langle\Delta v_z^{2}\rangle$ to $\sigma_z^{2}$ regardless of the stars' pre-existing motions, but at fixed energy the resulting jump shrinks on a warmer baseline, $\Delta\sigma_z\simeq\langle\Delta v_z^{2}\rangle/2\sigma_z$. Calibrating in $\Delta(\sigma_z^{2})$ rather than $\Delta\sigma_z$, the observed $4$~km~s$^{-1}$ jump on the Milky Way's $\sigma_z\simeq16$--$25$~km~s$^{-1}$ outer-disc baseline \citep{das+24} carries $1.2$--$1.8$ times the energy of the same jump in our colder disc, raising the inferred mass by a factor of $1.1$--$1.3$ -- within, and in the same direction as, the factor-$\sim1.5$ metric systematic above. A denser suite of simulations spanning the $(M, v_{\rm peri})$ plane could calibrate this inversion properly, making the fossil record a quantitative mass scale for ancient encounters.

\subsection{Fossil phase-space impacts from satellite interactions in other simulations}
\label{ssec:lit_comparison}

Other simulations have also examined fossil impacts of satellites on the phase-space structure of disc galaxies. \citet{quinn+93} and \citet{velazquez_white99} established that sinking satellites deposit their orbital energy in the disc, thickening and heating it permanently, and \citet{villalobos_helmi08,villalobos_helmi09} and \citet{kazantzidis+08} showed that mergers with mass ratios of order $5{:}1$ leave lasting thick-disc substructure and morphological signatures. At total mass ratios of order $1{:}10$ (Table~\ref{tab:sim_params}), our encounters are only moderately gentler in mass; the sharper distinction is that our satellite never merges, depositing only a fraction of its orbital energy in a single unbound passage rather than the entirety of it as its orbit decays.

The survival of such fossil records has also been examined in cosmological settings. \cite{martig+14a} and \cite{martig+14b} used a suite of seven simulated discs with a range of merger histories to identify which disc properties preserve a fossil record. Density profiles and scale heights of mono-age populations retain no trace of the merger history; the fossils instead lie in a jump in the age--velocity relation at the age of the event, imprinted by mass ratios as minor as 1:10 to 1:15, though with typical age uncertainties of order 30 per cent such jumps are substantially washed out, so the record can be intact in the disc and still be absent from the data. \citet{grand+16} reach similar conclusions across sixteen zoom-in simulations of Milky-Way-sized galaxies: the bar is usually the dominant driver of vertical disc heating, but the strongest individual heating events are external perturbations from satellites and subhaloes with masses above $\sim10^{10}$~M$_\odot$ -- a threshold all three of our satellites, at $0.7$--$1.1\times10^{11}$~M$_\odot$ (Table~\ref{tab:orbit_massloss}), comfortably exceed. 

\citet{thulasidharan+26} use TNG50 Milky-Way analogues to show that a galaxy's merger history is imprinted in the age--thickness relation, $\Delta_z({\rm age})$, closely analogous to our $|z|_{50}$(age) diagnostic. Their merger and flyby examples behave qualitatively differently, however: a $5{:}1$ merger produces a broad, step-like increase in $\Delta_z$ that heats pre-existing stars, whereas their flyby produces only a weak, narrow peak confined mainly to stars formed during the encounter from disturbed gas. Our flyby, by contrast, produces a sharp, permanent discontinuity in $|z|_{50}$ that affects pre-existing stars (Figure~\ref{fig:scale_height_vs_age}), not just newly-formed ones. \citet{grand+23} find in their cosmological simulation that a $\sim10^{10}$~M$_\odot$ satellite on a polar orbit at a similar lookback time of $\sim6$~Gyr leaves only a coherent, still-phase-mixing imprint -- a warp that winds into an `ever-present' phase spiral, weakest in stars older than the encounter -- with no reported heating of pre-existing stars. Our merger-like signature is therefore the outlier among simulated flybys but our satellites are nearly an order of magnitude more massive than either of these. 

A smaller number of studies simulate the Sagittarius--Milky Way interaction directly, though mostly focused on the transient, still-evolving response of the disc rather than its permanent record. \citet{purcell+11} showed that Sagittarius impacts can drive spiral structure and ring-like features in the outer disc, and \citet{gomez+13}, analysing the same simulations, that the passages excite vertical density waves. \citet{laporte+18} extended this work to include the Large Magellanic Cloud, finding that repeated passages flare and thicken the outer disc, and \citet{laporte+19,laporte+20,laporte+22} identified surviving fossils of these interactions in Gaia data, including kicked-up outer-disc material and the Anticentre Stream. The high-resolution suite of \citet{bennett+22} and the direct/indirect decomposition of \citet{asano+25} are chiefly concerned with the present-day, transient response. 

\subsection{Consistency of features in observed phase-space with an early Sagittarius encounter}
\label{ssec:obs_recovery}

\cite{mackereth+19} measure age-velocity dispersion relations across the Milky Way disc using APOGEE and Gaia, and find that the radial trend of $\sigma_z/\sigma_R$ with Galactocentric radius in the outer, low-$\alpha$ disc exceeds what disc-internal heating (GMCs, spiral arms) can produce, and tentatively attribute this radial excess to satellite-driven perturbation or disc warping. 

\cite{das+24} measure a jump in $\sigma_z$ of $\sim4$~km~s$^{-1}$ for $6$-Gyr-old stars beyond $\sim11$~kpc, against a $\sim1$~km~s$^{-1}$ measurement error; our own $\Delta\sigma_z=+4.1$~km~s$^{-1}$ for \texttt{pmr6}\ in the oldest age bin (Section~\ref{sec:sat_impacts}) matches this closely, and \texttt{pmr6\_massive}'s satellite mass ($1.06\times10^{11}$~M$_\odot$) and pericentre ($15.4$~kpc) closely match the heavier of the two degenerate solutions \citet{das+24} independently infer using the impulse approximation ($6\times10^{10}$~M$_\odot$ crossing at $12$~kpc, or $1\times10^{11}$~M$_\odot$ at $16$~kpc). This agreement should, however, be read as encouraging rather than as a clean validation: our simulated disc is substantially colder than the real Milky Way at every pre-passage age. Our host's $\sigma_z$ rises from only $\sim4.5$~km~s$^{-1}$ at a stellar age of $\sim0.5$~Gyr to $\sim9.3$~km~s$^{-1}$ just before the passage at $\sim5.5$~Gyr (Figure~\ref{fig:sigma_z_vs_age_heating}), whereas \citet{das+24} measure $\sigma_z\sim10$--$20$~km~s$^{-1}$ even for their youngest, least-heated outer-disc stars -- a factor of $2$--$3$ higher at comparable ages. The close match in $\Delta\sigma_z$ is therefore better understood as the encounter depositing roughly the right amount of energy, acting on a disc whose baseline dynamical state differs substantially from the real Galaxy's, rather than as evidence that our simulated disc reproduces the Milky Way's actual dynamical conditions.

\subsection{The plausibility of the proposed mass and orbit for Sagittarius}
\label{ssec:plausibility}

Dynamically-modelled estimates of Sagittarius' mass and orbit are most tightly constrained at recent times, when they can be calibrated against the present-day stream and bound remnant. \citet{vasiliev+21}'s fiducial model, for example, finds Sagittarius crossing the disc at $R\simeq24$--$25$~kpc with a mass of $\lesssim2\times10^{9}$~M$_\odot$ at a pericentric passage $\sim1.2$--$1.3$~Gyr ago. By their most recent modelled passage, $\sim40$~Myr ago, it had lost up to $90$~per cent of its initial dark matter mass and roughly half of its initial stellar mass. No equivalent dynamically-modelled mass or orbit estimate exists for the specific $5$--$7$~Gyr-ago encounter studied in this paper: constraints on that earlier epoch come only indirectly, via the stream and stellar-content evidence summarized in Section~\ref{sec:introduction}, rather than from a fitted orbit.

A pericentre of $R\simeq16$--$18$~kpc (Table~\ref{tab:orbit_massloss}), closer than Sagittarius' much more recent $\sim24$--$25$~kpc passage found by \citet{vasiliev+21} might at first seem implausible, since dynamical friction is generally expected to sink a satellite progressively closer to its host over time, not further away. However, \citet{santistevan+23} find in the FIRE-2 cosmological simulations that for $67$~per cent of satellites around Milky-Way-mass hosts that experience multiple pericentres, the most recent pericentre is not the minimum one reached over the satellite's history: the minimum is typically $\simeq37$~per cent smaller and occurs $\simeq6.3$~Gyr earlier. They identify no single dynamical origin, but rather two channels of comparable importance: roughly half of these satellites gradually gain specific angular momentum over time, which they attribute to a time-dependent and/or triaxial host halo potential, while the other half gain angular momentum in rapid episodes near a pericentre or apocentre, consistent with perturbations from other satellites and structure in the outer halo. The triaxiality channel finds direct support in the Sagittarius-specific modelling of \citet{vasiliev+21}, in which the apocentre decays through dynamical friction while the pericentre varies non-monotonically owing to the non-sphericity of the halo potential. 

If the early perturber was indeed as massive as $\sim0.6$--$1\times10^{11}$~M$_\odot$, satellite population statistics identify it almost uniquely. As argued by \citet{das+24}, a Milky-Way-mass host is expected to harbour only $\sim$one satellite with a total mass of $\sim10^{11}$~M$_\odot$ \citep{read+19,santos+22}, and the Milky Way already hosts two unusually massive satellites -- the LMC and Sagittarius, each with masses of up to $\sim10^{11}$~M$_\odot$ -- making the existence of a third, as-yet-undiscovered object of this mass very unlikely. Of the two, the LMC can be excluded for the early epoch as well as the present one. In the standard first-infall picture it was still far beyond the Milky Way's virial radius $5$--$7$~Gyr ago, and even in the second-passage scenario of \citet{vasiliev24}, its previous pericentre, $5$--$10$~Gyr ago, occurred at a distance $\gtrsim100$~kpc. This leaves Sagittarius as the only plausible candidate for the early encounter.

A further plausibility question is whether such an early encounter can reasonably be treated as impulsive. \citet{das+24} adopt exactly this approximation \citep{chew_wick52} to infer the mass and pericentre of the possible early perturber. Our satellites here are unbound by construction to simulate this impulsive scenario. A bound encounter would have been slower, but still possibly impulsive: a crossing at $\sim300$~km~s$^{-1}$ would deliver its kick over $t_{\rm enc}\sim110$~Myr, still shorter than the vertical oscillation periods beyond the crossing radius ($T_z\gtrsim190$~Myr at $R\gtrsim15$~kpc; Section~\ref{sec:orbit}), so the encounter would remain impulsive where the fossil resides, if less deeply so. E.g. \cite{hunt+21} finds that in their model it is specifically the first crossing that acts impulsively, with the later, slower and deeper passages coupling adiabatically instead. For the inference, assuming the $M/v_{\rm peri}$ degeneracy of Section~\ref{ssec:heating_scaling} still holds, a slower bound crossing would require a lighter satellite.

\subsection{Caveats and limitations}
\label{ssec:caveats}

Our experiment is deliberately idealized, and its relation to the real Galaxy's history should be kept in view. The simulations model an isolated Milky-Way-plus-satellite pair. There is no cosmological environment, no subsequent mergers or accretion events, and no Large Magellanic Cloud and each run contains a single pericentric passage from which the satellite departs unbound, whereas the real Sagittarius returned repeatedly \citep{ruiz_lara+20} and survives as a bound remnant today (Section~\ref{ssec:plausibility}). The real disc has therefore accumulated many perturbations since any early passage, none of which are captured here. These later perturbations erode the fossil's datability from the young side: the step survives only if post-encounter heating of younger stars remains below its $\sim2$--$4$~km~s$^{-1}$ amplitude. The later Sagittarius passages were slower, less impulsive (Section~\ref{ssec:plausibility}) and an order of magnitude less massive \citep{vasiliev+21}, so this is plausible, but it compounds the age-uncertainty washout of \citet{martig+14b} as a reason the observable record may understate the intrinsic one.

The fidelity of the model disc itself imposes further limits. The pre-existing bulge, thin-disc, thick-disc and halo populations are `painted' rather than self-consistently grown (Section~\ref{sec:simulation_setup}), so only the newly-formed thin-disc stars analysed in Section~\ref{sec:sat_impacts} are free of initial-condition artefacts, and nothing here constrains how the pre-existing populations would have responded to the passage. The newly-formed disc is also substantially colder than the real Milky Way at every pre-passage age (Section~\ref{ssec:obs_recovery}), so absolute amplitudes should be compared with observations with care, and our $R_{50}$, $|z|_{50}$ and $\sigma_z$ estimators are subject to resolution and particle-noise limits, particularly in the youngest age bins and the outermost radii, mitigated but not eliminated by the $\geq30$-star threshold and equal-number radial binning.

Numerical discreteness itself is quantifiable. Two-body scattering of disc stars by the $3\times10^{5}$~M$_\odot$ halo particles heats the disc spuriously \citep{lacey_ostriker85,ludlow+21}; from the measured halo density and dispersion at $R\simeq9$~kpc we estimate a contribution of $\sim20$~(km~s$^{-1}$)$^{2}$ to $\sigma_z^{2}$ over $10$~Gyr, or $10$--$20$~per cent of the host's secular heating budget, so the host's heating relation is partly numerical. This component is, however, shared identically by all four runs and subtracts out of every differential measurement. The one discreteness channel the control does not share -- the satellite's $23\times$ heavier dark-matter particles crossing the disc after the passage -- is measured to be negligible: fewer than $25$ such particles ($\lesssim3\times10^{8}$~M$_\odot$) lie within $30$~kpc of the host at any time after the crossing, and none at all in \texttt{pmr6\_faster}. Consistently, stars formed after the passage match the control to $\lesssim0.3$~km~s$^{-1}$ over the following $7$~Gyr, bounding any run-specific spurious heating at that level.

\section{Conclusions}
\label{sec:conclusions}

Motivated by evidence for an early encounter between the Milky Way and Sagittarius some $5$--$7$~Gyr ago, we have run four controlled $N$-body+hydrodynamical simulations of a Milky-Way-like host. There is a fiducial run in which a Sagittarius-like satellite makes a single, unbound, near-polar passage through the outer disc (\texttt{pmr6}), a run with a $1.5\times$ more massive satellite (\texttt{pmr6\_massive}), a run with a higher impact speed (\texttt{pmr6\_faster}), and an isolated-host control with no satellite at all (\texttt{pmr6\_host}). We analysed the present-day phase-space structure of the newly-formed thin-disc stars in each, using density and velocity dispersion profiles and associated metrics ($R_{50}$, $|z|_{50}$, $\sigma_z$) by age. Our main conclusions are as follows.

\begin{enumerate}
    \item The satellite makes a single pericentric passage at $r_{\rm peri}\simeq16$--$18$~kpc, crossing the disc almost perpendicularly and impulsively for the disc beyond the crossing radius, and leaves on an unbound trajectory. Mass loss is confined almost entirely to the passage itself, lasting only $\Delta t_{\rm strip}\simeq0.1$--$0.2$~Gyr. The stripped fraction is governed by the orbit rather than the satellite's mass: the fiducial and massive satellites lose nearly identical fractions ($\sim50$~per cent), while the faster satellite, which spends less time near pericentre, loses only $34$~per cent.

    \item The passage leaves a permanent, age-dependent fossil signature in the disc's vertical structure. Stars old enough to have existed at the encounter (ages $\gtrsim7.1$~Gyr) are permanently thickened relative to the isolated host. $|z|_{50}$ reaches $0.35$~kpc in \texttt{pmr6\_massive} against $0.26$~kpc in the host, a $\sim37$~per cent excess -- while stars that formed after the passage are indistinguishable from the control. The onset of the divergence coincides with the passage age in all three satellite runs.

    \item The same fossil appears in the kinematics: $\sigma_z$(age) departs from the host's smooth, internally-set age--velocity dispersion relation only above the passage age, reaching excesses over the host heating relation of $\Delta\sigma_z\simeq+5.4$, $+4.1$ and $+2.8$~km~s$^{-1}$ in \texttt{pmr6\_massive}, \texttt{pmr6}\ and \texttt{pmr6\_faster}\ respectively. In \texttt{pmr6\_massive}, the excess heating is also localized in radius, peaking near the satellite's disc-crossing radius.

    \item By contrast, the disc's radial structure carries no comparably clean fossil: $R_{50}$(age) shows the same inside-out growth trend in all four runs, with only \texttt{pmr6\_massive} showing a clear departure. 

    \item The simulated fossil is quantitatively consistent with the observed one: \texttt{pmr6}'s $\Delta\sigma_z=+4.1$~km~s$^{-1}$ matches the $\sim4$~km~s$^{-1}$ jump measured by \citet{das+24}, whose independently inferred perturber solutions ($0.6$--$1\times10^{11}$~M$_\odot$ at pericentres of $12$--$16$~kpc) include one, $1\times10^{11}$~M$_\odot$ at $16$~kpc, that essentially coincides with \texttt{pmr6\_massive}'s parameters. Our simulated disc is however $2$--$3\times$ dynamically colder than the real Milky Way at every pre-passage age, so this agreement is best read as the encounter depositing the right amount of energy rather than as a full validation.
\end{enumerate}

Several questions remain open. We still need to explore whether the heating is imparted directly by the satellite's tidal field or mediated by the host halo's collective response, as in the dark-matter-wake scenario of \citet{grand+23}. How the real Sagittarius, having survived its first passage, returned for the subsequent encounters recorded in the disc's star-formation history \citep{ruiz_lara+20} is not addressed by our single-flyby experiments. And whether the fossil signature identified here is recoverable in real data, where stellar age uncertainties of $\sim30$~per cent can wash out exactly this kind of discontinuity \citep{martig+14b}, deserves further study. Finally, the encounter modelled here should also leave chemical fossils in the age--metallicity structure of the outer disc and in the metal-poor gas delivered by the satellite. We will examine this in a companion paper.

\section*{Acknowledgements}
PD acknowledges the support of a UKRI Future Leaders Fellowship UKRI2762. PD also acknowledges the use of Claude Code Sonnet 5 and Opus 5 for generating code for analysing the simulations from pseudocode, rewriting the manuscript text in places, and finding and synthesising papers. All ideas, tests, and code structure are the authors' own. JH acknowledges the support of a UKRI Ernest Rutherford Fellowship ST/Z510245/1. PD and JH would also like to thank visiting PhD student Qikang Feng in his help during early explorations of the data.

\section*{Data Availability}
 
The simulations are available on reasonable request to the authors. The code generated to carry out the analysis and produce the figures will also be shared on reasonable request.



\bibliographystyle{mnras}
\bibliography{payeldas,jason} 


\appendix


\bsp	
\label{lastpage}
\end{document}